\documentclass[journal]{rmaa}

\usepackage{paralist}
\usepackage{psfrag,color}

\usepackage[latin1]{inputenc}
\usepackage[T1]{fontenc}
\def\Fe{[Fe/H]}
\def\fov{FoV}
\def\RRab{RRab}
\def\RRc{RRc}
\def\Tv{$t_V$}
\def\Ti{$t_I$}

\title{NGC 1261: A time-series \emph{VI} study of its variable stars}

\author{
A. Arellano Ferro,\altaffilmark{1}
I.H. Bustos Fierro,\altaffilmark{2} 
J.H. Calder\'on,\altaffilmark{2,3}
J.A. Ahumada,\altaffilmark{2}}

\altaffiltext{1}{Instituto de Astronom\'ia, Universidad Nacional Aut\'onoma de
M\'exico, Ciudad Universitaria, C.P. 04510, M\'exico.}	

\altaffiltext{2}{Universidad Nacional de C\'ordoba. Observatorio
Astron\'omico. C\'ordoba  Argentina.}

\altaffiltext{3}{Consejo Nacional de Investigaciones Cient\'ificas y
T\'ecnicas (CONICET), C\'ordoba, Argentina.}

\shortauthor{Arellano Ferro et al.}
\shorttitle{NGC 1261 \emph{VI} variable stars time-series}

\fulladdresses{
\item A. Arellano Ferro (armando@astro.unam.mx): Instituto de Astronom\'{i}a, Universidad Nacional Aut\'onoma de
M\'exico, M\'exico.

\item I Bustos Fierro (ivanbf@oac.unc.edu.ar), J. H. Calder\'on (calderon@oac.unc.edu.ar), J. A. Ahumada (javier.ahumada@unc.edu.ar): Universidad Nacional de C\'ordoba. Observatorio Astron\'omico. C\'ordoba Argentina.}

\listofauthors{A. Arellano Ferro, I.H. Bustos Fierro, J.H. Calder\'on, \& J.A. Ahumada}
\indexauthor{Bustos Fierro, I.H.}
\indexauthor{Calder\'on, J.H.}
\indexauthor{Ahumada, J.A.}

\abstract{Time-series \emph{VI} CCD photometry of the globular cluster NGC 1261 is employed 
to study its variable star population. A membership analysis of most
variables based on Gaia~DR2 proper motions and colours, was performed prior to
the estimation of the mean cluster distance and
metallicity. For the member RR Lyrae, their light curves were Fourier
decomposed to calculate their individual values of distance, [Fe/H], radius and mass. The $I$-band P-L for RR Lyrae stars was also employed. Our best
estimates of the metallicity and distance of this Oo I cluster are [Fe/H]$_{\rm ZW}$=$-1.42
\pm 0.05$ dex and $d=17.2 \pm 0.4$ kpc. No mixture of fundamental and first
overtone RR Lyrae stars in the either-or or bimodal region is seen in this cluster, as
it seems to be the rule for Oo~I clusters with a red horizontal branch. A multi-approach search in a region of about 10$\times$10 arcmin$^2$ around the
cluster  revealed no new variable stars within the limitations of our CCD photometry.}

\resumen{Empleamos una serie temporal de im\'agenes CCD
en las bandas \emph{VI} del c\'umulo globular NGC~1261, para estudiar su poblaci\'on de estrellas variables.
Se realiz\'o un an\'alisis de la membres\'{\i}a de las estrellas
variables basado en movimientos propios y colores dados en Gaia~DR2.  Por
medio de la descomposici\'on de Fourier de curvas de luz de
estrellas RR~Lyrae,
se obtuvieron sus valores individuales
de distancia, [Fe/H], radio y masa.
Tambi\'en utilizamos la relaci\'on
P-L en la banda \emph{I} para las RR~Lyrae.
Nuestras estimaci\'ones de la metalicidad y la distancia medias de
NGC~1261, un c\'umulo de tipo Oo~I, son [Fe/H]$_{\rm ZW} = -1.42\pm 0.05$~dex
y $d = 17.2\pm0.4$~kpc. No se observa mezcla de modos de pulsaci\'on fundamental y primer sobre tono de RR~Lyrae en la regi\'on bi-modal, como parece ser la norma
para c\'umulos de tipo~OoI con rama horizontal roja. Una b\'usqueda cuidadosa
en la regi\'on de $10'\times10'$ centrada en el c\'umulo no revel\'o
nuevas variables, dentro de las limitaciones de nuestra fotometr\'{\i}a.}

\addkeyword{globular clusters: individual: NGC 1261}
\addkeyword{stars: variables: RR Lyrae } 
\addkeyword{stars: fundamental parameters}

\begin{document}
\maketitle

\section{Introduction}

The southern globular cluster NGC 1261 (C0310$-$554), also known as Caldwell 87, is located in the constellation Horologium ($\alpha = 03^{h} 12^{m} 16.21^{s}$, $\delta$ = $-55^{\circ}$ 12' 58.4'', J2000, \citet{Goldsbury2010}). It was discovered by James Dunlop in 1826, and remained little studied until the mid-XXth century. It is a cluster of intermediate brightness, located far from the bulge of the Galaxy ($l$ = 270.54$^{\circ}$, $b$ = 55.12$^{\circ}$) and, consequently, has negligible or no reddening. The Catalog of Parameters for Milky Way Globular Clusters compiled by \citet{Harris1996}
 (2010 edition),  lists a metallicity [Fe/H]=$-1.27$, a distance to the Sun of 16.4 kpc, the level of the Horizontal Branch (HB) as $V_{\mathrm{HB}}=16.7$ mag,  and a reddening $E(B-V) = 0.01$. The Catalogue of Variable Stars in Globular Clusters (CVSGC), \citep{Clement2001}  (2017 edition), lists 31 stars, although only 29 have been confirmed as variables: RR Lyrae stars (23), SX Phe (3), variable red giants or SR (2) and eclipsing binaries EC? (1). The members of the RR Lyrae population were first reported from photographic studies by  \citet{Laborde1966} later on by \citet{Bartolini1971},  \citet{Wehlau1977a}, \citet{Wehlau1977b}  and, more recently, in the CCD studies of \citet{Salinas2007} and  \citet{Salinas2016}, who
  discovered five RR Lyrae stars, three SX Phoenicis, one long period variable and an eclipsing binary of the W UMa type.
  
  So far, the variable star population of NGC 1261 has not been studied with the specific aim of estimating some of the physical properties of the cluster, such as the mean metallicity, distance and age, by performing a detailed analysis of the light curves. Likewise, the distribution of variables in the HB has not been discussed, along with an analysis of individual membership.
  Studies of the distribution of RR Lyrae pulsation modes in the instability strip around the first overtone red edge (FORE), have shown that the modes are neatly splitted in all Oo II clusters with blue HB's; however,  this happens  only in some of the Oo I type clusters with redder HB's (e. g. \citet{Arellano2018b}). It has been argued that this property is related to the distribution of stars in the Zero-Age Horizontal Branch (ZAHB) which, in turn, depends on the mass loss during the He-flashes at the Asymptotic Giant Branch (AGB) (\citealt{Arellano2018a}; \citealt{Caputo1978}). The fact that NGC~1261 is among the few Oo I clusters with extremely red HB, has intermediate metallicity and harbours numerous RR  Lyrae stars of both modes, add a particular interest to its study.
  
  In the present work we perform such analysis based on a new, extensive time-series of $VI$ CCD images. The specific calibrations for diferent families of variable stars, concerning Fourier light curve decomposition, luminosity of the horizontal branch,
and P-L relations are discussed.
 We also make use of the possibility of ensuring the membership of the variables to the cluster through the analysis of proper motions available in the Gaia~DR2 catalogue. The paper is structured in the following way: In $\S 2$, we describe our observations, the data reduction process and the transformation to the standard photometric system. In $\S 3$, we report the periods, epochs and display the \emph{VI} light curves of all variables detected in our photometry. We discuss the cluster membership and some cases of clear light contamination by extremely close neighbours, as well as the corresponding corrections in amplitude and position, in the color-magnitude diagram (CMD). Our failed efforts to find new variables are described. In $\S 4$, we present the log $P$-amplitude diagram or Bailey's diagram to confirm the Oo I type nature of NGC 1261. $\S 5$ describes our estimation of the interstellar reddening. 
  $\S 6$ $\S 7$ and $\S 8$ contain the light curve Fourier decomposition and physical parameters estimations, and the comparison with previous determinations of [Fe/H] and distance. In $\S 9$ we comment on the structure of the HB and the correlation with the distribution of RR Lyrae pulsation modes in the inestability strip. In $\S 10$ we summarize our work.

\section{Data, observations and reductions}

The data used for the present work were obtained with the 1.54$-$meter telescope at the Bosque Alegre Astrophysical Station of the Cordoba Observatory, (National University of Cordoba), Argentina, during seventeen nights between August 19th 2017 and November 30th 2018. A total 308 and 388 $V$ and $I$ CCD images were acquired. 
The detector used was a CCD Alta F16M of 4096$\times$4096 square 9-micron pixels, binned $2\times2$, with a scale of 0.496 arcsec/pix (after binning). The images were trimmed to $1200\times1200$ pixels, for a useful field of view (FoV) of approximately $10\times10$ arcmin$^2$

Table \ref{tab:observations} summarizes the observation dates, exposure times and average seeing conditions.

\begin{table}
\scriptsize
\caption{The time distribution of $VI$ observations of NGC~1261.}
\centering
\begin{tabular}{lccccc}
\hline
Date (y/m/d)         &  $N_{V}$ & $t_{V}$ (s) & $N_{I}$ &$t_{I}$ (s)&Avg. seeing (") \\
\hline

2017/08/19 & 17  & 400      & 20  & 200  & 3.0\\
2017/08/20 & 19  & 400      & 25  & 200  & 2.9 \\
2017/09/10 & 15  & 400      & 17  & 200  & 3.1 \\
2017/09/15 & 15  & 400      & 15  & 200  & 3.0 \\
2017/09/22 & 24  & 400      & 29  & 200  & 2.3 \\
2017/10/06 & 15  & 400      & 19  & 200  & 3.0 \\
2017/10/28 & 26  & 400      & 30  & 200 & 3.1 \\
2017/12/07 & 13  & 400      & 14  & 200 & 2.4 \\
2018/08/03 & 4  & 300   & 6  & 150-200  & 2.9 \\
2018/08/04 & 27  & 300  & 33  & 150  & 3.4 \\
2018/08/05 & 27  & 300  & 27  & 150  & 2.9 \\
2018/08/11 & -- & --  & 23  & 150 & 3.0 \\
2018/09/02 & 33  & 300  & 37  & 150  & 2.4 \\
2018/09/14 & 28  & 300   & 38  & 150  & 2.3 \\
2018/09/16 & 23 & 300   & 29 & 150  & 2.3 \\
2018/11/16 & 18  & 300  & 21 & 150  & 2.9 \\
2018/11/30 & 4  & 300   & 5  & 150 & 2.6 \\
2018/12/16 & 22  & 300   & 24  & 150 & 2.5 \\

\hline
Total:   & 330  & -- & 412   & -- & --  \\
\hline
\end{tabular}
\label{tab:observations}
\raggedright
\center{Columns $N_{V}$ and $N_{I}$ give the number of images taken with the $V$ and $I$
filters respectively. Columns \Tv and \Ti provide the exposure time,
or range of exposure times employed during each night for each filter. The 
average seeing is listed in the last column.}
\end{table}

\subsection{Difference Image Analysis}

For the reduction of our data, we employed the software Difference Imaging Analysis (DIA) with its pipeline implementation DanDIA\footnote{DanDIA is built from the DanIDL library of IDL routines
available at \url{http://www.danidl.co.uk}} (\citealt{Bramich2008}; \citealt{Bramich2013}). With this, we were able to obtain high-precision photometry for all the point sources in the FoV of our CCD images. First, a reference image is created by DanDIA by stacking the best images in each filter; then, the reference image is subtracted from the rest of the images. 
In each reference image, we measured the fluxes (referred to as
reference fluxes) and positions of all PSF-like objects (stars) by
extracting a spatially variable empirical
PSF. This PSF is built from about 300-400 isolated stars, and a third degree polynomial is fitted to each detected object.

Differential fluxes are converted into total fluxes.  
The total flux $f_{\mbox{\scriptsize tot}}(t)$ in ADU/s at each epoch $t$ can be estimated as:
\begin{equation}
f_{\mbox{\scriptsize tot}}(t) = f_{\mbox{\scriptsize ref}} +
\frac{f_{\mbox{\scriptsize diff}}(t)}{p(t)},
\label{eqn:totflux}
\end{equation}

\noindent
where $f_{\mbox{\scriptsize ref}}$ is the reference flux (ADU/s), $f_{\mbox{\scriptsize diff}}(t)$ is the differential flux (ADU/s) and
$p(t)$ is the photometric scale factor (the integral of the kernel solution). Conversion to instrumental magnitudes was achieved using:
\begin{equation}
m_{\mbox{\scriptsize ins}}(t) = 25.0 - 2.5 \log \left[ f_{\mbox{\scriptsize tot}}(t)
\right],
\label{eqn:mag}
\end{equation}

\noindent
where $m_{\mbox{\scriptsize ins}}(t)$ is the instrumental magnitude of the star at time $t$. The above procedure has been described in detail in \citet{Bramich2011}. 

\begin{figure} 
\includegraphics[width=8.0cm,height=8.0cm]{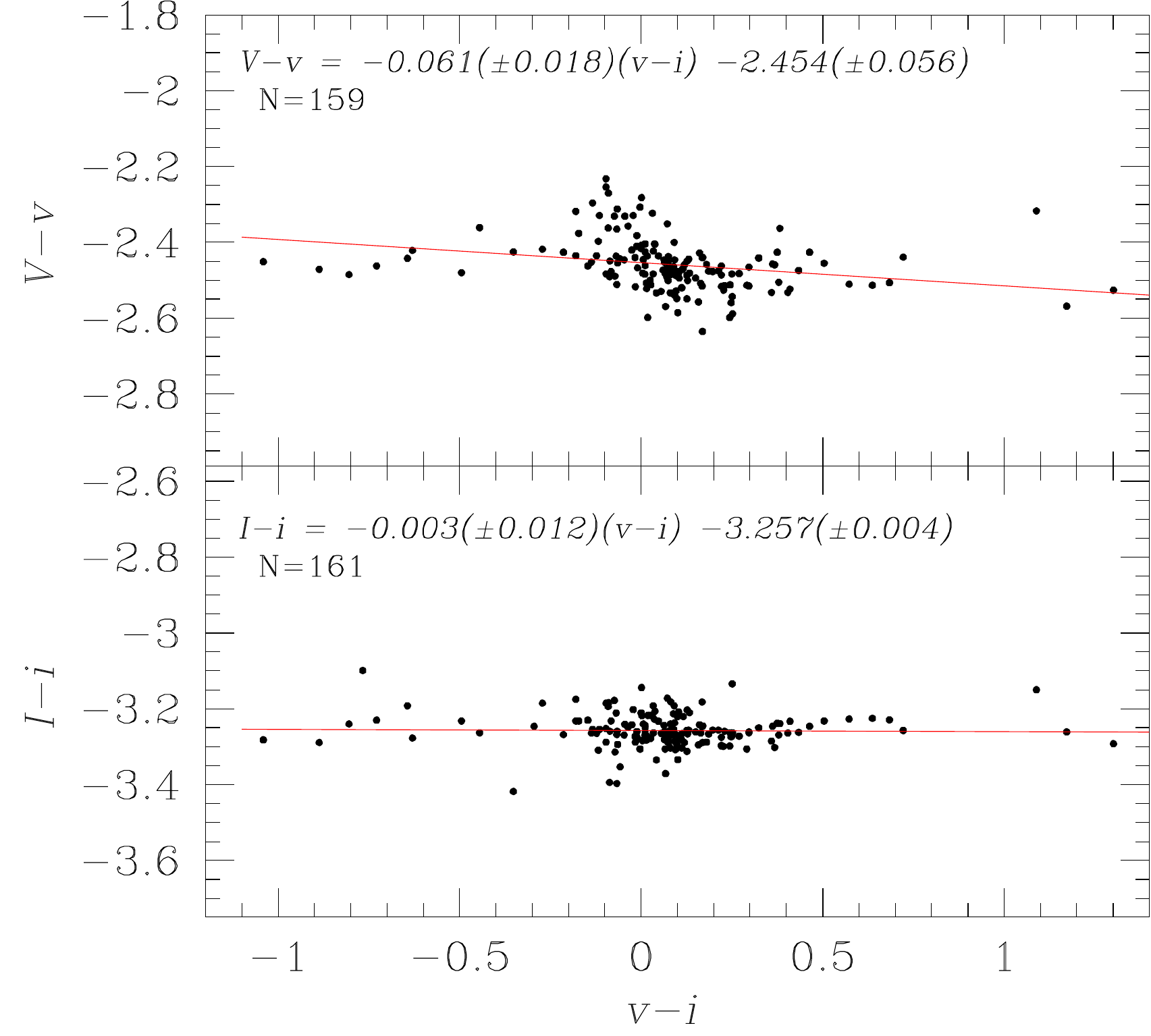}

\caption{Transformation relations obtained for the $V$ and $I$ filters between the instrumental and the standard photometric systems. We employed a set of standard stars in the field of NGC 1261 from the collection of \citet{Stetson2000}.}
    \label{transV}
\end{figure}
%

\subsection{Photometric Calibrations}

\subsubsection{Relative calibration}
\label{sec:rel}

To correct for possible systematic errors, we applied the methodology developed in \citet{Bramich2012} to solve for the magnitude offsets $Z_{k}$ that should be applied to each photometric measurement from the image $k$. In terms of DIA, this translates into a correction for the systematic error introduced into the photometry due to a possible error in the flux-magnitude conversion factor \citep{Bramich2015}. In the present case the corrections were very small, $\sim$ 0.001 mag for stars brighter than $ V \sim 18.0$.

\subsubsection{Absolute calibration}
\label{absolute}

Standard stars in the field of NGC 1261 are included in the online collection of \citet{Stetson2000}\footnote{\url{http://www3.cadc-ccda.hia-iha.nrc-cnrc.gc.ca/community/STETSON/standards}} and we used them to transform instrumental \emph{vi} magnitudes into the Johnson-Kron-Cousins standard \emph{VI} system. These stars are distributed in the cluster periphery, they are generally isolated and can be accurately measured by DanDIA. The mild colour dependence of the standard minus instrumental magnitudes is shown in Fig. \ref{transV}. The transformation equations are explicitly given in the figure itself.

\subsubsection{Internal errors}
\label{rms}

\begin{figure} 
\includegraphics[width=8.0cm,height=8.0cm]{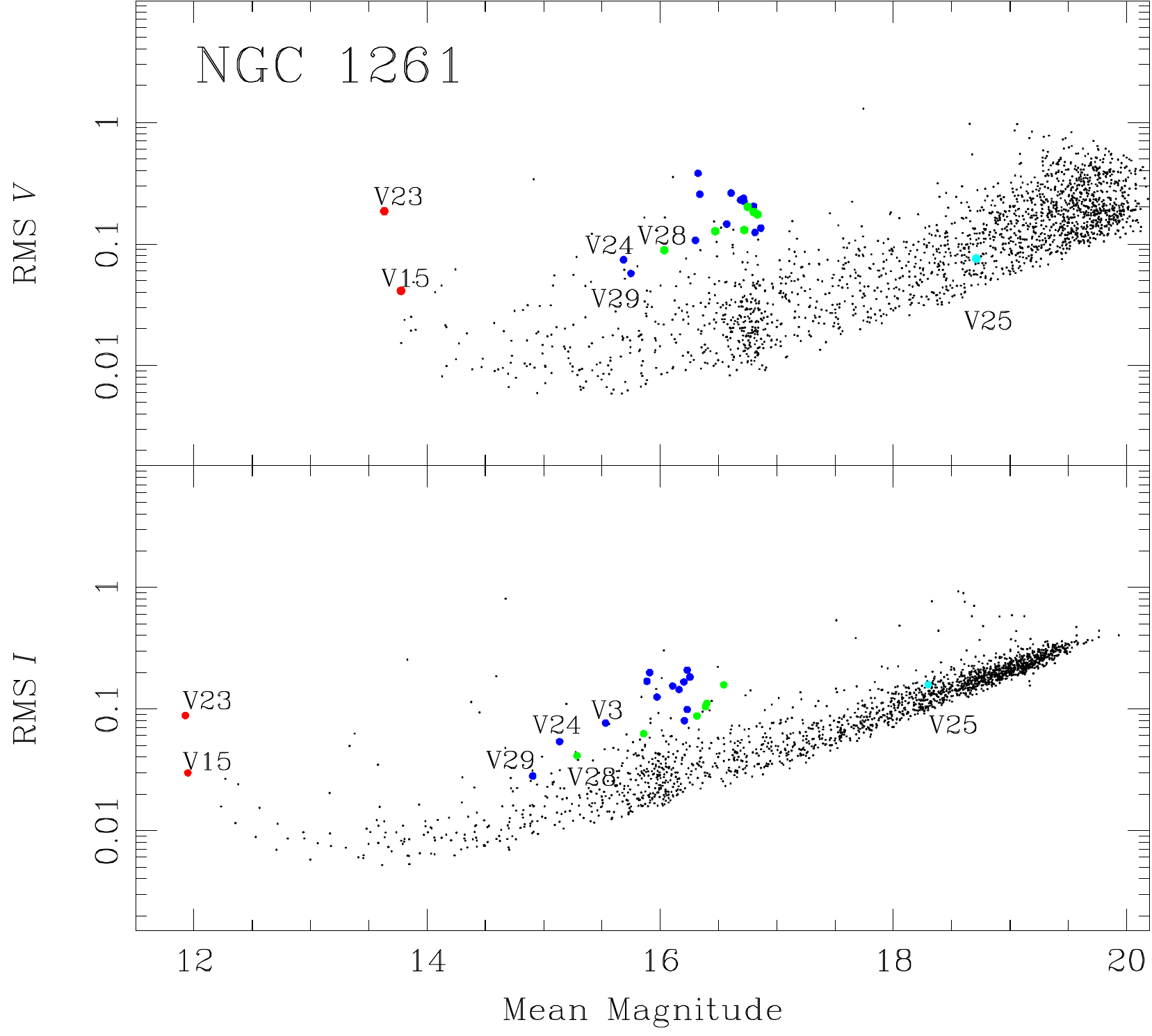}
\caption{rms of all light curves measured in our images as a function of the mean magnitude. Colour symbols indicate the position of the known variable stars as follows: blue and green are RRab and RRc stars respectively, red are SR stars and turquoise is the SX Phe star V25. }
\label{rmsVI}
\end{figure}

The internal errors of our CCD photometry can be evaluated via the rms diagram of Fig. \ref{rmsVI}.

\section{Variable stars in NGC 1261}

\begin{figure*} 
\includegraphics[width=18.0cm,height=14.0cm]{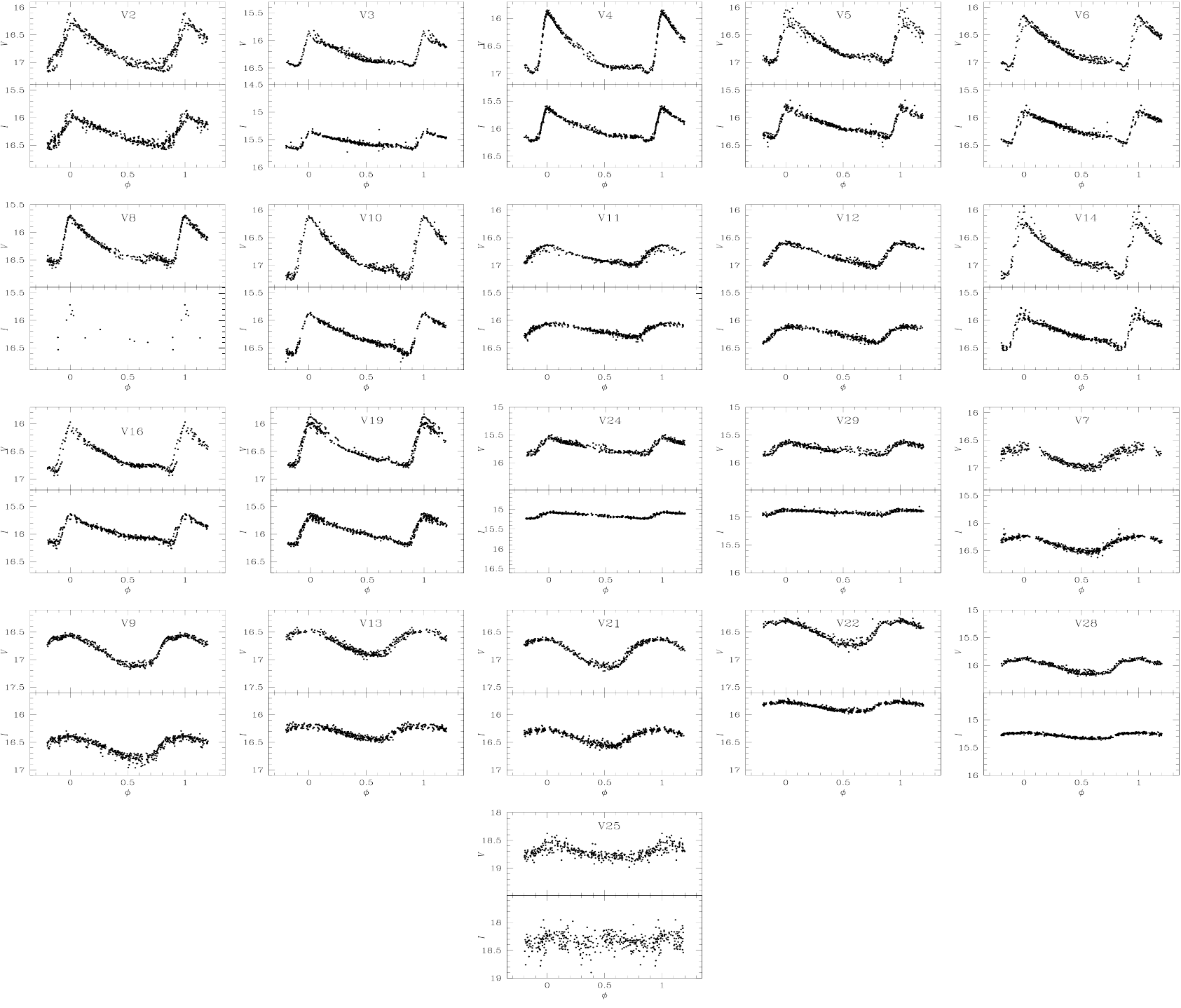}
\caption{$\emph{VI}$ light curves of known RRab, RRc stars and the SX Phe V25 in  NGC 1261.}
\label{N1261_RRlc}. 
\end{figure*}

There are 29 variable stars listed in the Catalogue of Variable Stars in Globular Clusters (CVSGC) \citep{Clement2001}. The time-series \textit{VI} photometry obtained in this work is reported in Table  \ref{tab:vi_phot}, of which only a small portion is included in the printed version of the paper. The full table shall be available in electronic form in the Centre de Donn\'es astronomiques de Strasbourg database (CDS). The RR Lyrae stars V20 and V27 are blended in our images and could not be resolved, therefore they are neither included in the Table \ref{tab:vi_phot} nor discussed in this paper. The  light curves in our data for the RRab, RRc and the SX Phe V25 are displayed in Fig. \ref{N1261_RRlc}. Our determination of their mean magnitudes, amplitudes and periods are given in Table \ref{variables}. The rms diagrams in the $V$ and $I$ filters and the Color-Magnitude diagram (CMD) of the cluster are shown in  Figs. \ref{rmsVI} and \ref{CMD} respectively. The positions of the known variable stars are shown.

\begin{table}
\scriptsize
\begin{center}
\caption{Time-series \textit{VI} photometry for the variables stars observed in this work$^*$}
\label{tab:vi_phot}
\centering
\begin{tabular}{cccccc}
\hline
Variable &Filter & HJD & $M_{\mbox{\scriptsize std}}$ &
$m_{\mbox{\scriptsize ins}}$
& $\sigma_{m}$ \\
Star ID  &    & (d) & (mag)     & (mag)   & (mag) \\
\hline
 V2 & $V$& 2457985.75380& 16.864&  19.293&  0.010 \\   
 V2 & $V$& 2457985.75851&  16.830&  19.259&  0.010 \\
\vdots   &  \vdots  & \vdots & \vdots & \vdots & \vdots  \\
 V2 & $I$ & 2457985.74544 & 16.417 & 19.673 & 0.019  \\  
 V2 & $I$ & 2457985.74783 & 16.367 & 19.622 & 0.018  \\ 
\vdots   &  \vdots  & \vdots & \vdots & \vdots & \vdots  \\
 V3 & $V$ & 2457985.75380 & 16.304 & 18.752 & 0.006 \\   
 V3 & $V$ & 2457985.75851 & 16.294 & 18.741 & 0.006 \\
\vdots   &  \vdots  & \vdots & \vdots & \vdots & \vdots  \\
 V3 & $I$ & 2457985.74544 & 15.535 & 18.791 & 0.009 \\    
 V3 & $I$ & 2457985.74783 & 15.537 & 18.794 & 0.009 \\   
\vdots   &  \vdots  & \vdots & \vdots & \vdots & \vdots  \\
\hline
\end{tabular}
\end{center}
* The standard and
instrumental magnitudes are listed in columns 4 and~5,
respectively, corresponding to the variable stars in column~1. Filter and epoch of
mid-exposure are listed in columns 2 and 3, respectively. The uncertainty on
$\mathrm{m}_\mathrm{ins}$ is listed in column~6, which also corresponds to the
uncertainty on $\mathrm{M}_\mathrm{std}$. A full version of this table is available at the CDS database.

\end{table}

It is rather clear that the scatter around the horizontal branch (HB) is large and that some RR Lyrae stars appear much too bright. It is common to use the star position on the CMD as an argument against membership in the cluster, e.g., RR Lyrae stars that appear much above or below the HB. The information now available in the Gaia~DR2 data base \citep{Gaia2018}, enables a deeper discussion on possible contaminations by neighbouring not-resolved stars and their proper motions. In the following section we shall refer to individual RR Lyrae stars with peculiar positions in the CMD.

Regarding the SX Phe stars, there are three of them reported in the
CVSGC: V25, V26 and V30.
Our light curve of V25 is shown in Fig.  \ref{variables}. However, we have been unable to detect the faint stars V26 and V30, near the cluster core, probably because of the seeing conditions during our observations. Nevertheless we want to point out that we see no stars, or that they are badly blended, in the positions marked in the identification chart of the discovery paper \citep{Salinas2007}.

\subsection{On the cluster membership and light contamination of individual variables}
\label{contam}

\begin{figure*} 
\includegraphics[width=17.0cm,height=10.0cm]{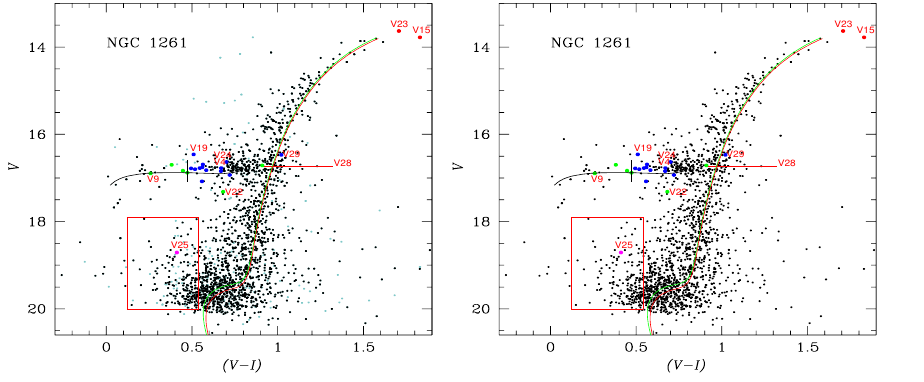}
\caption{Colour-magnitude diagram of NGC 1261. In the  left panel, black and light blue symbols are used to distinguish between cluster members and non-members. The right panel display only the cluster members. On the HB, blue and green symbols stand for RRab and RRc stars. Red and pink symbols represent SR and SX Phe stars. We searched for new SX Phe variables within the Blue Straggler region defined as the red box and found none. Isochrones and ZAHB are from the models of \citet{Vandenberg2014} for [Fe/H]=-1.4, Y=0.25 and [$\alpha$/H]=0.4 for 10 and 11 Gyr, and have been shifted  to a distance of 17.2 kpc, and a reddening $E(B-V)$=0.01. }
    \label{CMD}
\end{figure*}

With the aim of producing a Colour-Magnitude Diagram (CMD) free of field stars, we applied the method of \citet{Bustos2019}  to identify probable members in the
field of the cluster. The method uses the high quality astrometric data available in Gaia~DR2, and is
based on the Balanced Iterative Reducing and Clustering using Hierarchies (BIRCH) algorithm (Zhang
et al. 1996) in a four-dimensional space of physical
parameters --positions and proper motions-- that detects groups of stars in that 4D-space. We extracted
5258 Gaia sources that are very likely members of the
cluster. The membership analysis shows that most of
the stars in this field are indeed members of the cluster, but that
their proper motions are quite small ($\sim 2$ mas/yr), and that their distribution is well overlapped with that of the field stars. The top two panels of Fig. \ref{CMD} display the full CMD diagram before and after the field star correction. Almost all known variables are matched with a member. The exceptions are stars V24, V26 and V31, that match with
sources of unknown membership, because 
their proper motions were not measured in the Gaia survey.

Plotting the positions of all Gaia~DR2 sources in the field of our images, we noted that, in some cases, two or even three of them  can fall within the FWHM of the PSF of a detected stellar source. A direct consequence of this is an alteration of the magnitude of the star, making it apparently brighter and, in the case of variable stars, artificially reducing the amplitude of their light curves. 

\begin{figure*} 
\includegraphics[width=16.9cm,height=6.8cm]{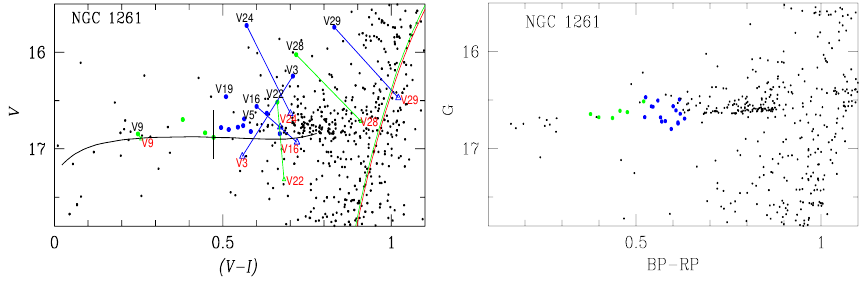}
\caption{Expansion of the HB branch of NGC 1261. 
The left panel shows the HB region built from the \emph{VI} indices of this work. Labelled stars are those with more than one Gaia~DR2 source within their PSF. Their corrected positions in the CMD are represented by open triangles and labelled with red numbers, see $\S$ \ref{contam} for details. The vertical black line is the empirical position of the red-edge first-overtone instability strip (FORE), defined by \citet{Arellano2016}. (See the discussion in $\S$ \ref{decomp} and $\S$ \ref{HB_STRUC}). The right panel shows the HB region built from the Gaia~DR2 photometric indices. With the exception of the peculair stars discussed in $\S$ \ref{contam} the resulting distribution of the RR Lyrae stars is comparable in both panels.}
    \label{HB_GAIA}
\end{figure*}

In Fig. \ref{HB_GAIA}, we show an amplification of the HB region of NGC~1261. In the left panel, known variables are plotted using the intensity weighted mean  $<V>$ and $<I>$ listed in Table \ref{variables}. RRab and RRc stars are represented by blue and green symbols respectively. It is evident that some of the RR Lyrae stars are much too bright relative to the ZAHB, namely, V3, V19, V22, V24, V28 and V29. The positions of the two RRc stars V22 and V28 are also much redder than expected. As a reference, the vertical black line that represents the empirical red edge of the first overtone instability strip (FORE) is also displayed \citep{Arellano2016}.

To help decide whether the odd positions of these stars are the result
of light pollution, or simply because they are non-members, in the cluster,
we analyzed the photometric values of all Gaia sources present in a
given variable star PSF. The magnitudes of the individual sources
in the Gaia photometric system, $G$-, $G_\mathrm{BP}$-, and $G_\mathrm{RP}$-band,
were
transformed into $V$ and $I$ magnitudes using the relationships
provided by J.M.~Carrasco (2018: Gaia team), and available in Section 5.3.7 of the Gaia DR2
documentation\footnote{\url{http://gea.esac.esa.int/arcfhive/documentation/GDR2/index.html}}. Then, the expected combined magnitude was calculated, and hence
  an estimation was derived of the correction to be applied to our observed (combined) $V$
  and $I$ magnitudes. This also enabled us to estimate the amplitude corrections.
 
\begin{figure} 
\includegraphics[width=7.0cm,height=11.0cm]{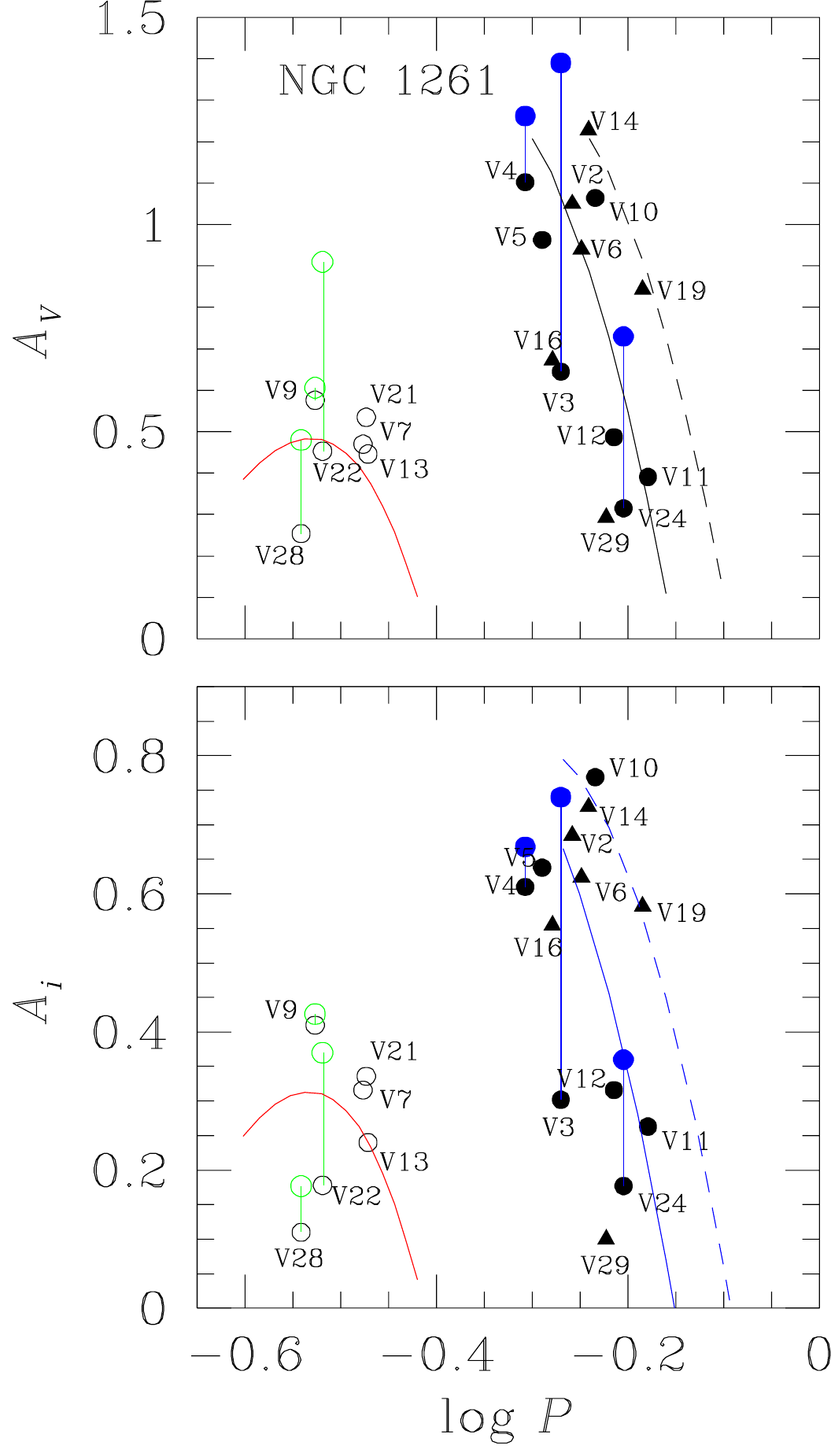}
\caption{Period-Amplitude diagram of NGC 1261. The amplitudes of a few variables have been corrected by the presence of an unresolved neighbour. These are plotted with green and blue symbols for RRc and RRab stars respectively. Triangles are used for Blazhko variables. The distribution of stars is characteristic of Oo I type clusters. See $\S$ \ref{contam} and $\S$ \ref{baileyDiagram} for details.}
   \label{Bailey}
\end{figure}

In Table \ref{pollution} we list the variable stars with two Gaia sources within their PSF. We use the variability flag in Gaia DR2 to
identify the variable component when it is available; for stars without variability flag (V17, V24 and V29) we have
analysed their positions in the CMD with Gaia DR2 magnitudes and colours to decide which is the variable
and which is the contaminant.
Their Gaia magnitudes were transformed into the Johnson-Kron-Cousins photometric system ($V_{Gaia}$ , $I_{Gaia}$
and $V_{Gaia} - I_{Gaia}$), and the combined magnitudes of the pair, $V_{mix}$ and $I_{mix}$, were calculated. 
The differences between the magnitude of the variable component and the combined
magnitude of the pair, $\Delta V$ and $\Delta I$, were computed;
the corrected positions are represented by triangles in the left panel of Fig. \ref{HB_GAIA} and labelled with red numbers. The corrected amplitudes due to the presence of the neighbour star (Amp $V$ and  Amp $I$) were also calculated, and are represented by colour symbols and vertical displacements in the Bailey diagram, discussed below in $\S$ \ref{Bailey}. It can be seen in that diagram that,
after the corrections of the amplitudes, most stars have moved much closer to the expected locus for this OoI type cluster. 

The persistent peculiar positions of some variables in the CMD, make us suspicious about their membership status in the cluster, or that a serious light pollution by neighbouring stars is corrupting our photometry for these specific objects. In the right panel of Fig. \ref{HB_GAIA}, we reproduced the HB of the cluster using the Gaia~DR2 photometric indices. It is rewarding to see that all RR Lyrae stars fall properly on the HB, which also indicates that the selection of the true variable, performed in Table \ref{pollution}, for those stars contaminated by a secondary Gaia source, was correct. This figure also shows that the distribution of modes on the HB is neatly segregated, a property that will be discussed later in the paper.

We can conclude that, all known variables are most likely, cluster members, and that our photometry for these
stars, namely V3, V16, V19, V24, V28 and V29  is seriously contaminated by a close, well within the PSF, unseen neighbour. 

We note that the two SR V15 and V23, are mem-
bers while for the three SX Phe, V25 lies in the blue
straggler region and seems a likely member, V26 has
not values of proper motion and V30 is definitely not
a member. The latter two were not detected in our
photometry.  Hence, our calculations of the physi-
cal parameters will be restricted to the non-peculiar
stars and will be described below.

\begin{table*}
\scriptsize
\begin{center}
\caption{Data of variable stars in NGC 1261 in the \fov~ of our images.}
\label{variables}

\begin{tabular}{llccccccccc}
\hline
Star ID & Type & $<V>$ & $<I>$   & $A_V$  & $A_I$ & $P$    &  $\rm HJD_{\rm max}$ & $\alpha$ (J2000.0)  & $\delta$  (J2000.0)   &  Gaia~DR2 ID    \\
        &      & (mag) & (mag)   & (mag)  & (mag) & (days) &  + 2450000 &                     &         \\
\hline
V2 \emph{Bl} & RRab & 16.780 & 12.285 & 1.16 & 0.76 &  0.585730  & 8095.6967& 03:12:11.28&$-$55:12:22.0 & 4733794790512123904\\
V3  & RRab  & 16.248 & 15.540 & 0.65 & 0.30& 0.537003   & 8439.8438   & 03:12:21.78&$-$55:13:50.7 & 4733793764014756224\\
V4  & RRab  & 16.633 &  16.002 &  1.10 & 0.61 & 0.492876     & 8376.7150   & 03:12:18.56&$-$55:13:28.4  & 4733794519928245888\\
V5  & RRab  & 16.690 & 16.127 & 0.96 & 0.64 & 0.513313  & 8012.7930 & 03:12:11.93&$-$55:13:01.9 & 4733794588651585664\\
V6 \emph{Bl}  & RRab   & 16.759 & 16.199 & 0.94 & 0.62 & 0.564100    & 8335.8557     & 03:12:25.05&$-$55:13:08.6 & 4733793867093968896\\
V7  & RRc  & 16.835 & 16.388 & 0.47 & 0.32 & 0.333546    & 8095.5598    & 03:11:58.44&$-$55:10:37.3 & 4733797852825688704\\
V8&RRab&16.283&--&0.86&--&0.538204&8012.8146&03:12:00.44&$-$55:15:16.4  & 4733700580405300480\\
V9  & RRc  & 16.846 & 16.598 & 0.58 & 0.41 & 0.297221  & 8019.6966     & 03:12:20.40&$-$55:13:35.3& 4733794519928956800\\
V10 & RRab  & 16.800 & 16.283 & 1.06 &0.77 & 0.583374 & 8364.7012     & 03:12:21.99&$-$55:11:45.9& 4733794726089585664\\ 
V11  & RRab & 16.843 &  16.174 &0.39& 0.26 & 0.662171 &8335.9157 & 03:12:05.50&$-$55:11:27.7& 4733794859231637376\\
V12  & RRab & 16.821 &  16.238 &0.49 & 0.32 & 0.610285 &8095.6920   & 03:12:26.15&$-$55:12:45.8 & 4733794657370122880\\
V13  & RRc  & 16.698 & 16.317 & 0.45 &0.24 &  0.337568& 8019.7891    & 03:12:07.01&$-$55:14:33.1 & 4733700713548568832\\
V14 \emph{Bl}  & RRab  & 16.775 &  16.228 &1.23 & 0.73&  0.573977& 8378.7791 & 03:12:09.75&$-$55:14:07.6 & 4733700717844254080\\
V15  & SR  & 13.778 & 11.948 &-- &-- & --  &8007.7702   & 03:12:02.50&$-$55:10:48.1& 4733797814168962304\\
V16 \emph{Bl} & RRab  & 16.561 & 15.961 & 0.67 & 0.56&   0.526160 &8469.6225 &03:12:13.90&$-$55:13:12.9 &4733794588651555584\\
V17$^{1}$&RRab&--&--&--&--&0.511631&7986.8406&03:12:15.50    &$-$55:12:36.2&4733794588651549824\\
V19 \emph{Bl} & RRab & 16.461 &  15.951 &0.84& 0.58 & 0.653738 &8376.7652            & 03:12:18.19&$-$55:12:44.7 & 4733794691727354624\\
V20$^{1}$&RRab&--&--&--&--&--&--&03:12:19.31&$-$55:13:00.4 & 4733794623011123200\\
V21  & RRc  & 16.881 & 16.408 & 0.54 & 0.34 &  0.336180 & 8439.8209     & 03:12:14.47&$-$55:12:31.3 & 4733794588651163648\\
V22  & RRc  & 16.516 & 15.855 & 0.45 & 0.18 & 0.302567  & 8033.8565    & 03:12:16.49&$-$55:13:38.1 & 4733794519931924224\\
V23  & SR  & 13.632 & 11.926 & -- & -- & --   & 8055.6833   & 03:12:15.69&$-$55:12:40.7 & 4733794588651583616\\
V24 & RRab  & 15.724 & 15.153 &0.32 &0.18&  0.624395&8376.6983  & 03:12:14.43&$-$55:13:34.8 & 4733794554291694592\\
V25 & SX Phe  & 18.710 & 18.300 &0.25 &0.18 &0.056503 &8033.7103  & 03:12:17.17&$-$55:11:22.7  & 4733794932246020480\\
V26$^{1}$ & SX Phe  & -- & -- &-- &--& -- &--     & 03:12:17.05&$-$55:12:43.9& 4733794588651476224\\
V27$^{1}$&SR&--&--&--&--&--&--&03:12:14.63    &$-$55:13:06.2&4733794588651284992\\
V28 & RRc  & 16.025 & 15.285 & 0.25 & 0.11 &  0.287349& 7986.7667 &03:12:13.53&$-$55:13:00.8 &4733794588651177344\\
V29 \emph{Bl} & RRab  & 15.742 & 14.911 &0.29 &0.10 & 0.598775&7986.8960  & 03:12:13.05&$-$55:13:20.5 & 4733794554291844352\\
V30$^{1}$ &  SX Phe  & --    & -- &-- & -- & --   & -- & 03:12:16.58&$-$55:12:54.0& 4733794588648149248\\ 
V31&  EC  & --    & -- & -- & 0.30 & 0.052493    & 8439.8246     & 03:12:18.70&$-$55:14:16.0&  4733794519931172608\\ 
\hline
\end{tabular}
\raggedright
\center{\quad \emph{Bl}: RR Lyrae with Blazhko effect.\\
1. Star not resolved in our images. }
\end{center}
\end{table*}

\subsection{Searching for new variables}

 Since prolonged time-series have proven successful in identifying new variables, we have performed a systematic search for them with the few strategies discussed below.

 On the CMD of NGC 1261, we isolated all stars contained  in regions where it is common to find variable stars, e.g. near the Instability Strip (IS) in the Horizontal Branch (HB), the Blue Straggler region (BS) and near the tip of the Red Giant Branch (TRGB). We analysed the light curves of the stars in those regions and looked for variability by determining their period (if any), using the program \textsc{period04} \citep{Lenz2005} or the string-length method (\citealt{Burke1970}; \citealt{Dworetsky1983}); then, we plotted their apparent magnitudes with respect to their phase. For long-term variables, the magnitude is plotted as a function of Heliocentric Julian Day in search for hints of variability This procedure recovered all known variables, but in all cases we found no new ones in our data.\\

A third method consists in the detection of variations of PSF-like peaks in stacked residual images, from which we can see the variable stars blink. Again, all previous known variables were detected, but no new variables emerged.

\subsection{About the RR Lyrae stars  from Gaia~DR2 in the field of NGC 1261}
\label{Clementini}

\begin{figure*} 
\includegraphics[width=17.0cm,height=8.5cm]{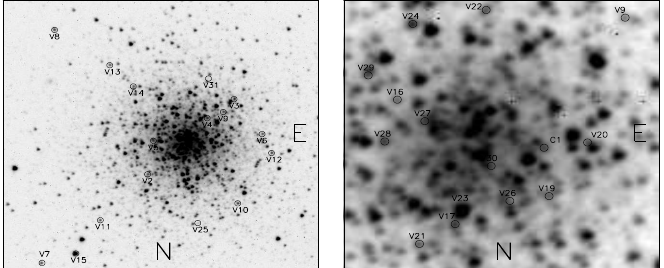}
\caption{Identification charts of all known variables in NGC 1261. The field of the left chart is 5.4$\times$5.4 arcmin$^2$, and the cluster core on the right chart is 1.3$\times$1.3 arcmin$^2$. The star labelled C1 is the one listed as an RR Lyrae by \citet{Clementini2019}, but that in this work is found to be non-variable (cf. $\S$ \ref{Clementini}). } 
    \label{MAPAS}
\end{figure*}

In their catalogue of RR Lyrae stars identified via the Specific Object Study pipeline in Gaia~DR2 all over the sky, \citet{Clementini2019} report 21 RR Lyrae stars within a radius of 20 arcmin around NGC 1261. Eighteen of these are known RR Lyrae stars listed in the CVSGC. Two stars, well in the outskirts of that field, (Gaia~DR2 sources  4734551739843543808 and 4733801701116515200), are not members according to the method of \citet{Bustos2019}  and are not in the field of our observations. A third star near the cluster core (Gaia~DR2 source 4733794519931744000), categorised by \citet{Clementini2019} as an RRc star, is present in our light curve collection. We have identified it as C1 in the chart of Fig. \ref{MAPAS}, but we found no variations. The star, with mean magnitudes $V$=16.35 and $I$=15.63, lies a bit above the red clump. Thus we do not confirm it as an RRc star.

For clarification, we offer in Fig. \ref{MAPAS} an identification chart of all variables reported in the CVSGC. The SX Phe stars V26 and V30, although identified according to the coordinates in the CVSGC, were not detected as variables in the present study.

\begin{table*}
\scriptsize
\begin{center}
\caption{Variable stars with two Gaia~DR2 sources within their PSF. For every pair, the Gaia DR2 source in the first line is the variable and the one in
the second line is the contaminant. See $\S$ 3.1 for details.}
\label{pollution}

\begin{tabular}{llccccccccccc}
\hline
Variable & Gaia DR2 ID & $\mathrm{G}$ & $G_\mathrm{BP} - G_\mathrm{RP}$   & $V_\mathrm{Gaia}$  & $I_\mathrm{Gaia}$ & $V_\mathrm{Gaia}-I_\mathrm{Gaia}$   &  $V_{mix}$ & $\Delta V$  & $I_{mix}$   &  $\Delta I$  & Amp $V$ & Amp $I$ \\
 &  &(mag) & (mag)   & (mag)  & (mag) & (mag) &  (mag)& (mag) & (mag)& (mag)&(mag)&(mag)   \\
  
\hline
V3& 4733793764014756224& 16.6773& 0.5234& 16.746& 16.295& 0.451& 15.91& 0.83& 15.32& 0.98& 1.39& 0.74\\
V3& 4733793764014757248& 16.4500& 0.8248& 16.591& 15.883& 0.708&--&--&--&--&--&--\\
\hline
V4& 4733794519928245888& 16.5663& 0.5417& 16.638& 16.172& 0.467& 16.50& 0.14& 16.07& 0.10& 1.26& 0.67\\
V4& 4733794519931768192& 18.7558&--& 18.773& 18.735& 0.038&--&--&--&--&--&--\\
\hline
V9& 4733794519928956800& 16.6788&  0.3993& 16.727& 16.377& 0.350& 16.67& 0.05& 16.34& 0.04& 0.61& 0.43\\
V9& 4733794519928957056& 19.9313&-- & 19.949& 19.910& 0.038&--&--&--&--&--&--\\
\hline
V16& 4733794588651555584& 16.5656&  0.6003& 16.650& 16.134& 0.516& 16.28& 0.37& 15.88& 0.25&--&--\\
V16& 4733794588651555840& 17.6072&--& 17.625& 17.586& 0.038&--&--&--&--&--&--\\
\hline
V17& 4733794588651549824& 16.6079& 0.6079& 16.694& 16.172& 0.522&15.76&0.93&15.14&1.03&--&--\\
V17& 4733794588651550080& 16.2219& 0.7993& 16.356& 15.670& 0.686& --& --& --& --&--&--\\
\hline
V22& 4733794519931924224& 16.6257& 0.4764& 16.686& 16.273& 0.413& 15.89& 0.80& 15.49& 0.78& 0.91& 0.37\\
V22& 4733794519931924352& 16.5920& 0.6806& 16.592& 16.217& 0.375&--&--&--&--&--&--\\
\hline
V24& 4733794554291694592& 16.4950& 0.6173& 16.583& 16.053& 0.530&15.67&0.91&15.27&0.78&0.73&0.36\\
V24& 4733794554291623424& 16.2512& 0.3297& 16.290& 15.996& 0.294& --& --& --& --& --& --\\
\hline

V27& 4733794588651284992& 16.5018& 0.5269& 16.571& 16.117& 0.454& 16.33& 0.24& 15.94& 0.17&--&--\\
V27& 4733794588651285376& 18.0489&--& 18.067& 18.028& 0.038&--&--&--&--&--&--\\
\hline
V28& 4733794588651177344& 16.5151& 0.5204& 16.583& 16.134& 0.449& 15.89& 0.69& 15.62& 0.52& 0.48& 0.18\\
V28& 4733794588651552256& 16.6861&--& 16.704& 16.665& 0.038&--&--&--&--&--&--\\
\hline
V29& 4733794554291844352& 16.5704&0.5458& 16.588& 16.550& 0.038& 15.92& 0.73& 15.64& 0.54&--&--\\
V29& 4733794554291843840& 16.6775&--& 16.695& 16.657& 0.038&--&--&--&--&--&--\\
\hline
\end{tabular}
\end{center}
\end{table*}

\section{Bailey diagram and Oosterhoff type}
\label{baileyDiagram}

The period-amplitude or Bailey diagram for RR Lyrae stars is shown in Fig. \ref{Bailey} for the \emph{VI} bandpasses. The periods and amplitudes are listed in Table \ref{variables}. In most cases, we took the amplitudes corresponding to the best fit provided by the Fourier decomposition of the light curves. In cases where the light curve showed Blazhko effect, the maximum amplitude was measured and the star was plotted with a triangular marker. The continuous and dashed black lines in the top panel of  Fig. \ref{Bailey} are
the loci for unevolved and evolved stars according to \citet{Cacciari2005}.
The red parabolas were calculated by \citet{Arellano2015} from a sample of RRc stars in five OoI clusters, avoiding Blazhko variables. In the bottom panel, the blue solid and segmented loci for unevolved and
evolved stars, respectively, are from \citet{Kunder2013a}.

As noted in $\S$ \ref{contam}, a few variables are clearly unresolved from very close neighbours. In these cases, the amplitudes were corrected using the individual magnitudes of the components listed in Table \ref{pollution}. The corrected amplitudes are indicated in Fig. \ref{Bailey} by green and blue symbols, for RRc and RRab stars respectively. It is rather clear that the amplitude  corrections help sustain the fact that the distribution of stars concentrates around the unevolved sequences, as expected in Oo I type clusters.\\

\section{Comment on the cluster reddening}
\label{REDDENING}
While this cluster is far away from the Galactic disk, and a very low reddening has been assigned to it, it is always a good exercise to calculate such reddening by an independent method. Individual reddenings for RRab stars can be estimated using the colour curve near minimum. Proposed originally by \citet{Sturch1966}, the method has now been calibrated in the $\emph{VI}$ passbands by \citet{Guldenschuh2005}, who concluded that the intrinsic colour $(V-I)_0$ of RRab stars curves  at phases 0.5-0.8, is 0.58 $\pm$ 0.02. We calculated the observed $(V-I)$ at this phase range in our light curves and estimated individual values of $E(V-I)$, the average of which, converted to $E(V-I)/1.259$ is  $0.055 \pm 0.051$, i.e., nearly zero, and in agreement with the interstellar reddening of $E(B-V)=0.01$ given by the calibrations of  \citet{Schlafly2011} and    \citet{Schlegel1998}. In what follows we shall adopt $E(B-V)=0.01$.

\section{RR Lyrae stars: \Fe and $M_V$ from light curve Fourier decomposition}
\label{decomp}

Fourier decomposition of the light curves and the use of empirical calibrations
enable us to estimate some key stellar physical parameters. The Fourier series to mathematically describe the light curve is of the form: 

\begin{equation}
    m(t) = A_{0} + \sum_{k=1}^{N}A_{k}\cos(\frac{2\pi}{P}k(t-E_{0}) + \phi_{k})
	\label{eq:quadratic}
\end{equation}

\noindent
where $m(t)$ is the magnitude at time $t$, $P$ is the period of pulsation, and $E_0$ is the epoch. When calculating the Fourier parameters, we used a least-squares approach to estimate the best fit for the amplitudes $A_k$ and phases $\phi_k$ of the light curve components.  The phases and amplitudes of the harmonics in Eq. \ref{eq:quadratic}, i.e. the Fourier parameters, are defined as $\phi_{ij} = j\phi_{i} - i\phi_{j}$, and $R_{ij} = A_{i}/A_{j}$. 

This approach has been
regularly used in RR Lyrae stars in a large number of clusters (e.g., \citet{Arellano2018a}; \citet{Deras2019}  and references therein). To avoid a lengthy
repetition, we refer the reader to Section 4 of \citet{Deras2019} for the specific
calibrations. Particularly, their Eqs. 4 and 5  describe the calibrations for RRab
stars of \citet{Jurcsik1996}  and \citet{Kovacs2001},  which render values of [Fe/H] and $M_V$, with standard deviations of 0.14 dex and 0.04 mag rectively, and to Eqs. 6 and 7, that give the calibrations for RRc stars of \citet{Morgan2007}  and \citet{Kovacs1998} with standard deviations of 0.14 dex and 0.042 mag respectively.
We have not included in the calculation of physical parameters those stars that are
apparently contaminated by a neighbouring source (cf. $\S$ \ref{contam}); since the presence of a neighbour also perturbs the light curves shapes, the Fourier parameters and the derived physical parameters turn out spurious.
In Table \ref{tab:fourier_coeffs}  we list the Fourier
parameters, and in Table \ref{fisicos} the
corresponding individual physical parameters for the stars included in the calculation and the averages. For comparison, we have transformed [Fe/H]$_{\rm ZW}$ on the \citet{Zinn1984} metallicity scale into the UVES scale using the equation [Fe/H]$_{\rm UVES}$= $-0.413$ + 0.130~[Fe/H]$_{\rm ZW} - 0.356$~[Fe/H]$_{\rm ZW}^2$ \citep{Carretta2009}. It should be noted that
the uncertainties we have quoted in Table \ref{fisicos} only represent the internal errors associated to the Fourier fitting procedure, and do not include any systematic errors
that may be inherent to the use of the calibrations quoted above to estimate the physical parameters. The standard deviation of the mean $\sigma_{\bar x}$, given below the weighted mean physical parameters, is comparable to the standard deviation of the calibrations, and represents a more reliable estimate of the systematic errors in our procedure.

\begin{table*}
\footnotesize
\caption{Fourier coefficients $A_{k}$ for $k=0,1,2,3,4$, and phases $\phi_{21}$, $\phi_{31}$ and $\phi_{41}$, for \RRab~  and \RRc~ stars. The numbers in parentheses indicate the uncertainty on the last decimal places.}
\centering                   
\begin{tabular}{lllllllll}
\hline
Variable ID     & $A_{0}$    & $A_{1}$   & $A_{2}$   & $A_{3}$   & $A_{4}$   &
$\phi_{21}$ & $\phi_{31}$ & $\phi_{41}$  \\
     & ($V$ mag)  & ($V$ mag)  &  ($V$ mag) & ($V$ mag)& ($V$ mag) & & & \\
\hline
       &       &   &   &   & RRab star    & &       &      \\
\hline
V2 & 16.788(4)& 0.331(6)& 0.126(6)& 0.072(6)& 0.029(6)& 4.036(59)& 8.198(99)& 6.147(219)\\
V6&  16.752(2)& 0.305(4) &0.146(4)& 0.098(3)& 0.070(4)& 3.910(33)& 8.215(51)& 6.205(68)\\   
V10& 16.810(3)& 0.377(4)& 0.189(3)& 0.128(4)& 0.077(4)& 3.936(27)& 8.255  (39)& 6.191(59)\\
V11& 16.837(3)& 0.147(4)& 0.060(3)& 0.028(4)& 0.011(4)& 4.050(78)& 8.587 (139)& 7.709(344)\\
V12&16.816(2)& 0.177(3)&0.070(3)& 0.034(3)& 0.018(3)& 4.406(48)& 9.254(90)& 7.522(150)\\    
V14&16.775(3)& 0.368(5)&0.182(4)& 0.154(5)& 0.088(5)& 4.079(39)& 8.369  (51)& 6.378(75)\\
\hline
             &                    &           &           & RRc stars &           
 &             &             &\\
\hline
V7& 16.810(3)& 0.181(4)& 0.011(4)& 0.019(4)& 0.008(4)& 4.665(31)&4.415(201)& 2.409(509)\\
V9&  16.846(1)& 0.285(2)& 0.052(2)& 0.028(2)& 0.028(2)&4.749(37)&2.734(64)& 1.485(67)\\
V13& 16.698(2)& 0.222(3)& 0.026(3)& 0.013(3)& 0.008(2)& 4.792(114)& 3.815(206)& 2.681(318)\\
V21& 16.881(1)& 0.262(2)& 0.022(2)& 0.023(2)& 0.006(2)& 4.795(88)& 4.115(88)& 2.729(347) \\

\hline	
\hline
\end{tabular}
\label{tab:fourier_coeffs}
\end{table*}

\begin{table*}
\footnotesize
\begin{center}
\caption[] {\small Physical parameters obtained from the Fourier fit for the \RRab~ and \RRc~ stars. The numbers in
parentheses indicate the uncertainty on the last decimal places. See $\S$ 6 for a detailed discussion.}
\label{fisicos}
\hspace{0.01cm}
 \begin{tabular}{llllllll}
\hline 
 &  &  & RRab star &  & & \\
\hline
Star&[Fe/H]$_{\rm ZW}$&[Fe/H]$_{\rm UVES}$ & $M_V$ & log~$T_{\rm eff}$  & log$(L/{L_{\odot}})$ & $M/{M_{\odot}}$&$R/{R_{\odot}}$\\

\hline
V2 &$-1.46(9)$ &$-1.37(10)$& 0.570(9)& 3.811(25) & 1.672(3) & 0.70(21)& 5.49(2)\\
V6 &$-1.49(5) $&$-1.40(5)$& 0.604(5)& 3.809(10) & 1.658(3) & 0.67(8)& 5.46(1)\\
V10 &$-1.53(4)$ &$-1.44(4)$ & 0.517(6)& 3.810(10) & 1.693(2) & 0.70(8)& 5.66(2)\\
V11 &$-1.51(13)$ &$-1.43(14)$ &0.600(6)& 3.790(39) & 1.660(2) & 0.67(31)& 5.98(2)\\
V12 &$-0.69(9)^{1} $&$-0.67(4)^{1}$& 0.635(4)& 3.813(18) & 1.646(2) & 0.55(12)& 5.29(1)\\
V14 &$-1.39(5)$ &$-1.28(5)$& 0.562(7)& 3.812(11) & 1.675(3) & 0.66(9)& 5.49(2)\\

\hline
Weighted Mean&$-1.42$&$-1.20$&0.590&3.810& 1.664&0.66 & 5.51\\
$\sigma_{\bar x}$ &$\pm$ 0.05  &$\pm$ 0.05 & $\pm$ 0.042 &$\pm$ 0.008 &$\pm$ 0.006& $\pm$ 0.05 & $\pm$ 0.09 \\
\hline

 &  &  & RRc stars &  & & \\
 \hline
Star&[Fe/H]$_{\rm ZW}$&[Fe/H]$_{\rm UVES}$ & $M_V$ & log~$T_{\rm eff}$  & log$(L/{L_{\odot}})$ & $M/{ M_{\odot}}$&$R/{R_{\odot}}$\\
\hline
V7 &$-1.06(42)$ &$-0.95(32)$ &0.571(18) &3.869(1) &1.672(7) &0.49(1)  &4.21(4)   \\
V9 &$-1.63(11)$ &$-1.57(13)$ &0.512(9) &3.866(1) &1.695(4) &0.63(1)  &4.37(2)  \\
V13  &$-1.50(41)$ &$-1.41(44)$ &0.588(10) &3.864(1) &1.677(4) &0.51(1) &4.47(2)  \\
V21  &$-1.31(18)$ &$-1.19(17)$ &0.571(10) & 3.866(1)&1.672(4) & 0.49(1)& 4.25(2) \\

\hline
Weighted Mean&$-1.51$&$-1.38$&0.547&3.866& 1.681&0.54 & 4.31\\
$\sigma_{\bar x}$  &$\pm$ 0.29  &$\pm$ 0.29 & $\pm$ 0.014 &$\pm$ 0.001 &$\pm$ 0.005& $\pm$ 0.03 & $\pm$ 0.05 \\
\hline
\end{tabular}
\end{center}
\center{
1. Not included in the average of [Fe/H].\\
}
\end{table*}

\section{Previous metallicity estimates of NGC 1261}

To our knowledge, no spectroscopic determination of the metallicity of NGC 1261 or of any of its stars has ever been published. Estimations of [Fe/H] based on photometric indices and their calibrations do exist; they use the 
height above the HB, the slope of the HB, and the intrinsic colour $(B-V)_0$ of the reg-giant branch (RGB) at the HB level,
\citet{Ferraro1993} employed these methods and found --1.54$\pm$0.4, --1.85$\pm$0.2 and --1.26$\pm$0.04 respectively, adopting a weighted average $-1.4\pm0.2$.
The value of [Fe/H] listed in the catalogue of \citet{Harris1996} (2010 edition) is $-1.27$, that corresponds to the compiled average of \citet{Carretta2009} in their UVES scale, which is equivalent to $-1.4\pm0.2$ in the Zinn-West scale.

Our calculations based on the Fourier decomposition and calibrations discussed in $\S$ \ref{decomp} and listed in Table \ref{fisicos} are, for the RRab stars: [Fe/H]$_{\rm zw} = -1.42 \pm 0.05$ or [Fe/H]$_{\rm UVES} = -1.20 \pm 0.05$, and for the RRc stars: [Fe/H]$_{\rm zw} = -1.51 \pm 0.09$ or [Fe/H]$_{\rm UVES} = -1.38 \pm 0.10$.  Since different empirical
relations have been used to estimate the metallicities of the
RRab and RRc stars, there may be some systematic offset between
the metallicity estimates for the two types of variable. We should note the peculiarly large value of [Fe/H] found for star V12. The reason for this is the anomalously large value of the Fourier parameter $\phi_{31}$ of 9.254$\pm$0.09. Star V12 is among the smallest amplitude and largest period RRab in NGC 1261, and reminds the case of star V12 in NGC 6171, whose light curve structure is different from that of other RRab stars \citep{Clement1997}, and then is not suitable for the calculation of [Fe/H] from the Fourier analysis. Although the associated uncertainties in our average values of [Fe/H] are smaller than in previous determinations, these values are slightly more metal-weak than those published; still, they are rather in agreement, within the given uncertainties. 

 \section{On the distance to NGC 1261}

The solar distance to NGC 1261 recorded in the catalogue of \citet{Harris1996} (2010 edition) is 16.3 kpc, calculated from the estimated mean $V = 16.70 \pm 0.05$ level HB of \citet{Ferraro1993}, adopting a [Fe/H]-$M_V$ relation with [Fe/H] and $E(B-V)=0.01$. Our Fourier estimates of  $M_V$ and $<V>$ for RRab and RRc stars lead to mean distances of $17.2\pm0.4$ and $17.6\pm0.7$ kpc respectively. Considering the large scatter of the 6 RRc stars in the CMD, and that we only used the RRab closer to the ZAHB, we should probably ignore the distance obtained from the RRc stars, in which case our best estimate of the distance to the cluster from this method is $17.2\pm0.7$ kpc. In Fig. \ref{CMD} we have shifted the ZAHB and isochrones to this distance, although, admittedly an eye fit of the HB suggests a brighter ZAHB for a distance of about 16.8 kpc 

Another independent method to estimate the distance to the cluster, is via the P-L relations of SX Phe stars and of RR Lyrae stars. In the former case, we have calculated $M_V$ of the star V25, the only SX Phe detected in our study, using three versions of P-L relations, from \citet{CohenSara2012}, \citet{Poretti2008} and \citet{Arellano2011}, and found 16.1, 15.5 and 15.0 kpc respectively, i.e., all values well below the estimates based on the HB luminosity.

Alternatively, the P-L in the $I$-band for RR Lyrae stars derived by \citet{Catelan2004}; $M_I = 0.471-1.132~ {\rm log}~P +0.205~ {\rm log}~Z$, with ${\rm log}~Z = [M/H]-1.765$; $[M/H] = \rm{[Fe/H]} - \rm {log} (0.638~f + 0.362)$ and log~f = [$\alpha$/Fe] \citep{Sal93}, was applied to the RRab and RRc stars residing close to the ZAHB in the CMD of Fig. \ref{CMD}. From 6 RRab and 4 RRc stars we found an average distance of $16.65 \pm 0.27$ kpc, which agrees very well with the results obtained from the Fourier light curve decomposition approach. 
In summary, the best results for the cluster distance are obtained from the Fourier
light curve decomposition of the RRab stars that show little scatter near the ZAHB, and the P-L ($I$) relationship; these methods give an average of $16.7$ kpc and an uncertainty between 0.7
and 0.3 kpc, respectively.

\section{Structure of the HB and Age}
\label{HB_STRUC}

\begin{figure} 
\includegraphics[width=7.9cm,height=7.9cm]{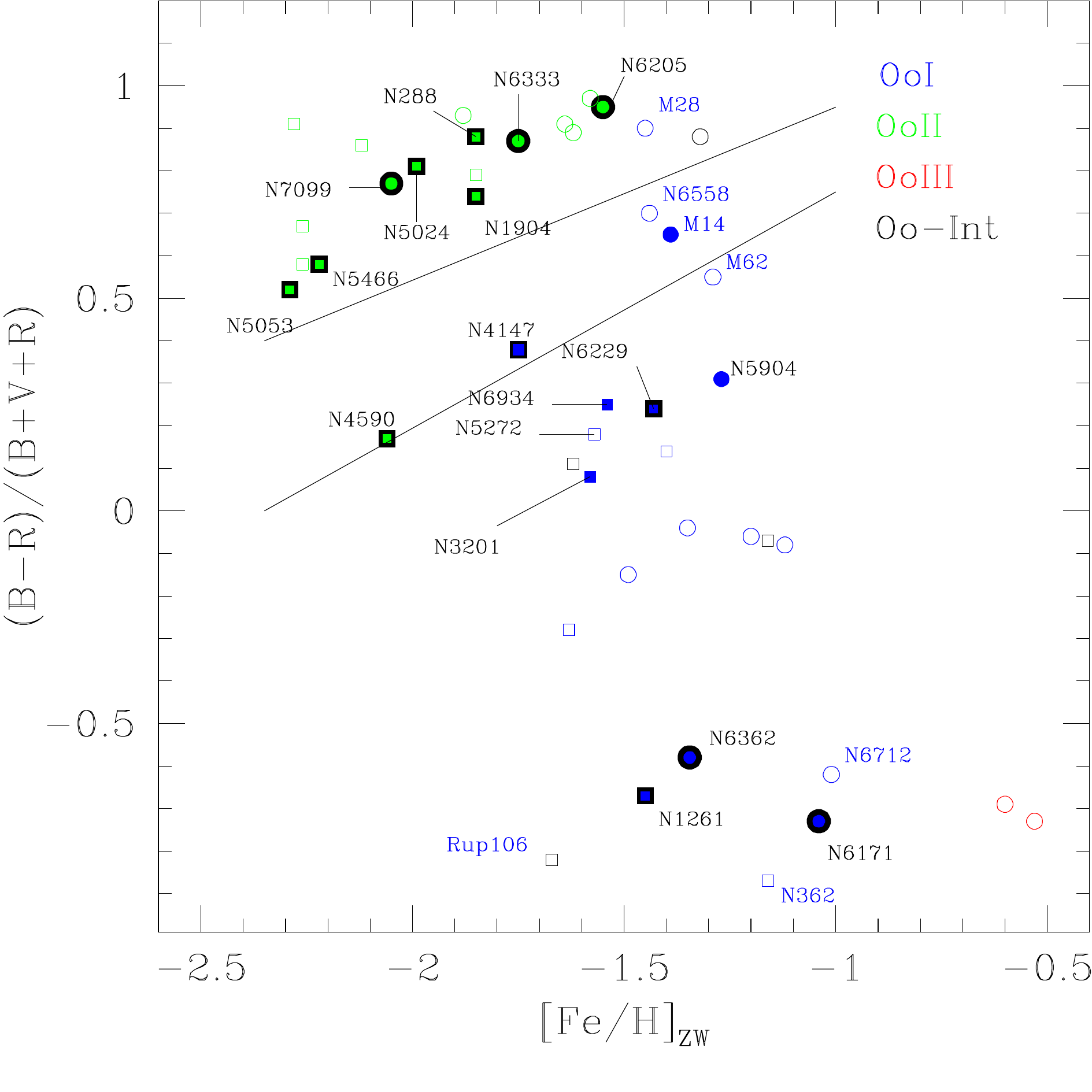}
\caption{The HB structure parameter $\mathcal L$ as function of [Fe/H]. Circles and squares are used for inner and outer halo clusters respectively. The black-rimmed symbols represent globular clusters where the fundamental
and first overtone modes are well segregated around the FORE, as opposed to filled non-rimmed symbols. The upper and lower solid lines are the limits of the Oosterhoff
gap according to \citet{Bono1994}.
Empty symbols are clusters not yet studied by our group.} 
    \label{CATELAN}
\end{figure}

NGC 1261 has a very red HB. The HB structure can be characterised  by the parameter $\mathcal L$ $\equiv (B-R)/(B+V+R) = -0.67$.  In the HB region shown in the left panel of Fig. \ref{HB_GAIA}, built from our \emph{VI} indices, the RRab and RRc stars appear clearly well segregated by the FORE. This is confirmed in the right panel, built with the Gaia DR2 indices. Fig. \ref{CATELAN} is an updated version of Fig. 9 of \citet{Arellano2018b} which includes NGC 1261 and a few other clusters.
Clearly NGC 1261, like other Oo I clusters with very red HBs like NGC~6171 (M107) and NGC~6362, shows a neat segregation of pulsation modes among the RR Lyrae stars. Other Oo I clusters with a similarly red HB where the RRc-RRab distribution is yet to be explored are: NGC 362, Rup 106 and NGC 6712. In Rup 106, however, no RRc stars are known, and  NGC 6712 is an interesting  borderline case, since either a few RRab are sitting in  the either-or region, or the FORE is shifted to the red by only about 0.04 mag (Deras 2019; private communication), which is the uncertainty in the location of the FORE. 
We have found that, in all Oo II clusters, the modes are always segregated. Only a few Oo I clusters with $\mathcal L$ between 0.0 and 0.7 have been found to have inter-mode region populated by RRab and RRc stars: NGC 3201, NGC 5904 (M5), NGC 6402 (M14) and NGC 6934.

 Based on Hubble Space Telescope Advanced Camera for Surveys photometry, and using the isochrone fitting between the base of turnoff and the lower RGB, \citet{Vandenberg2013} concluded that the age of NGC 1261 is 10.75 $\pm$ 0.25 Gyr. Being
the precision of our photometry, and the dispersion in the CMD near the TO, inadecuate for a subtle discussion on the age of the system, we overplotted to our observed CMD two isochrones for 10.0 and 11.0 Gyr from the $Victoria-Regina$ stellar models of \citet{Vandenberg2014}, for [Fe/H]=$-1.4$, Y=0.25 and [$\alpha$/Fe]= 0.4. The isochrones and the ZAHB, shifted to a distance of 17.2 kpc and $E(B-V) = 0.01$. We can see (Fig. \ref{CMD}) that the current data set is consistent with the age of the cluster determined by \citet{Vandenberg2013}. 

\section{Summary and Discussion}

Almost all photometric studies of NGC~1261, both in the photographic plate and CCD eras, dedicate their efforts to the overall study of the cluster, its CMD structure and the estimation of its age. Only a few studies, since the pioneering works of \citet{Fourcade1966} and \citet{Laborde1966} when the first variables in this cluster were discovered, have been devoted to the study of the cluster variable stars population. Specifically in the CCD era, and to the best of our knowledge, only the works of \citet{Salinas2007} and \citet{Salinas2016} have focused on the study of the variable stars in crowded environments, via the differential image approach (DIA). Yet these authors point out the extreme difficulty of transforming the relative fluxes rendered by DIA, into the astrophysically more useful magnitudes in a given calibrated system.
We must add that time-series do not always cover the light curves fully in the phase diagrams. The present study is the first one, as far as we know,
that reports complete light curves in the \emph{VI} filters, and use them to derive physical quantities from the variable star families of NGC 1261. This approach also allowed our best estimate of the cluster distance to be $17.2\pm0.4$ kpc. Individual radii and masses for the RR Lyrae stars are also reported.

A thorough multi-approach search in a region of about $10\times10$ arcmin$^2$ around the cluster  revealed no new variable stars within the limitations of our CCD photometry. The Fourier decomposition technique was employed to derive the mean values [Fe/H]$_{\rm zw} = -1.45 \pm 0.02$ and [Fe/H]$_{\rm zw} = -1.51 \pm 0.09$ for the RRab and RRc stars, respectivelly. These two estimations should be considered independent, as they come from totally unrelated calibrations. The above results show a cluster slightly more metal poor and distant than generally accepted (e.g., \citealt{Harris1996}).

The distribution of cluster member RR Lyrae stars on the HB shows a rather clear splitting of modes about the FORE, as seems to be the case of other two or perhaps three globulars with very red HB, namely, NGC~6171, NGC~6362 and probably NGC~6712. Two more clusters with a very red HB are NGC~362 and Rup 106, but no detailed study of the distribution of modes exists for NGC~362,  and no RRc stars are known in Rup 106.

\section*{Acknowledgements}

We are grateful to Cecilia Qui\~nones for her proficient help during the acquisition of the observations in Bosque Alegre. This project was partially supported by DGAPA-UNAM (Mexico) via grant IN106615-17. We have made extensive use of the SIMBAD, ADS services.







\bibliographystyle{rmaa}
\bibliography{1261_RMAA_2}

\begin{thebibliography}
\expandafter\ifx\csname natexlab\endcsname\relax\def\natexlab#1{#1}\fi
\expandafter\ifx\csname href\endcsname\relax
  \def\href#1#2{}\fi
\expandafter\ifx\csname urllinklabel\endcsname\relax
  \def\urllinklabel{[LINK]}\fi
\expandafter\ifx\csname adsurllinklabel\endcsname\relax
  \def\adsurllinklabel{[ADS]}\fi

\bibitem[{{Arellano Ferro} {et~al.}(2018a){Arellano Ferro}, {Ahumada}, {Bustos
  Fierro}, {Calder{\'o}n}, \& {Morrell}}]{Arellano2018a}
{Arellano Ferro}, A., {Ahumada}, J.~A., {Bustos Fierro}, I.~H., {Calder{\'o}n},
  J.~H., \& {Morrell}, N.~I. 2018a, Astronomische Nachrichten, 339, 183


\bibitem[{{Arellano Ferro} {et~al.}(2011){Arellano Ferro}, {Figuera Jaimes},
  {Giridhar}, {Bramich}, {Hern{\'a}ndez Santisteban}, \&
  {Kuppuswamy}}]{Arellano2011}
{Arellano Ferro}, A., {Figuera Jaimes}, R., {Giridhar}, S., {Bramich}, D.~M.,
  {Hern{\'a}ndez Santisteban}, J.~V., \& {Kuppuswamy}, K. 2011, \mnras, 416,
  2265


\bibitem[{{Arellano Ferro} {et~al.}(2016){Arellano Ferro}, {Luna}, {Bramich},
  {Giridhar}, {Ahumada}, \& {Muneer}}]{Arellano2016}
{Arellano Ferro}, A., {Luna}, A., {Bramich}, D.~M., {Giridhar}, S., {Ahumada},
  J.~A., \& {Muneer}, S. 2016, \apss, 361, 175


\bibitem[{{Arellano Ferro} {et~al.}(2015){Arellano Ferro}, {Mancera Pi{\~n}a},
  {Bramich}, {Giridhar}, {Ahumada}, {Kains}, \& {Kuppuswamy}}]{Arellano2015}
{Arellano Ferro}, A., {Mancera Pi{\~n}a}, P.~E., {Bramich}, D.~M., {Giridhar},
  S., {Ahumada}, J.~A., {Kains}, N., \& {Kuppuswamy}, K. 2015, \mnras, 452, 727


\bibitem[{{Arellano Ferro} {et~al.}(2018b){Arellano Ferro}, {Rojas Galindo},
  {Muneer}, \& {Giridhar}}]{Arellano2018b}
{Arellano Ferro}, A., {Rojas Galindo}, F.~C., {Muneer}, S., \& {Giridhar}, S.
  2018b, Rev.Mex.Astron.Astrofis, 54, 325


\bibitem[{{Bartolini} {et~al.}(1971){Bartolini}, {Grilli}, \&
  {Robertson}}]{Bartolini1971}
{Bartolini}, C., {Grilli}, F., \& {Robertson}, J.~W. 1971, Information Bulletin
  on Variable Stars, 594


\bibitem[{{Bono} {et~al.}(1994){Bono}, {Caputo}, \& {Stellingwerf}}]{Bono1994}
{Bono}, G., {Caputo}, F., \& {Stellingwerf}, R.~F. 1994, \apj, 423, 294


\bibitem[{{Bramich}(2008)}]{Bramich2008}
{Bramich}, D.~M. 2008, \mnras, 386, L77


\bibitem[{{Bramich} {et~al.}(2015){Bramich}, {Bachelet}, {Alsubai}, {Mislis},
  \& {Parley}}]{Bramich2015}
{Bramich}, D.~M., {Bachelet}, E., {Alsubai}, K.~A., {Mislis}, D., \& {Parley},
  N. 2015, A\&A, 577, A108


\bibitem[{{Bramich} {et~al.}(2011){Bramich}, {Figuera Jaimes}, {Giridhar}, \&
  {Arellano Ferro}}]{Bramich2011}
{Bramich}, D.~M., {Figuera Jaimes}, R., {Giridhar}, S., \& {Arellano Ferro}, A.
  2011, MNRAS, 413, 1275


\bibitem[{{Bramich} \& {Freudling}(2012)}]{Bramich2012}
{Bramich}, D.~M. \& {Freudling}, W. 2012, MNRAS, 424, 1584


\bibitem[{{Bramich} {et~al.}(2013){Bramich}, {Horne}, {Albrow}, {Tsapras},
  {Snodgrass}, {Street}, {Hundertmark}, {Kains}, {Arellano Ferro}, {Figuera},
  \& {Giridhar}}]{Bramich2013}
{Bramich}, D.~M., {Horne}, K., {Albrow}, M.~D., {Tsapras}, Y., {Snodgrass}, C.,
  {Street}, R.~A., {Hundertmark}, M., {Kains}, N., {Arellano Ferro}, A.,
  {Figuera}, J.~R., \& {Giridhar}, S. 2013, \mnras, 428, 2275


\bibitem[{{Burke} {et~al.}(1970){Burke}, {Rolland}, \& {Boy}}]{Burke1970}
{Burke}, Edward~W., J., {Rolland}, W.~W., \& {Boy}, W.~R. 1970, Journal of the
  Royal Astronomical Society of Canada, 64, 353


\bibitem[{{Bustos Fierro} \& {Calder{\'o}n}(2019)}]{Bustos2019}
{Bustos Fierro}, I.~H. \& {Calder{\'o}n}, J.~H. 2019, \mnras, 488, 3024


\bibitem[{{Cacciari} {et~al.}(2005){Cacciari}, {Corwin}, \&
  {Carney}}]{Cacciari2005}
{Cacciari}, C., {Corwin}, T.~M., \& {Carney}, B.~W. 2005, \aj, 129, 267


\bibitem[{{Caputo} {et~al.}(1978){Caputo}, {Castellani}, \&
  {Tornambe}}]{Caputo1978}
{Caputo}, F., {Castellani}, V., \& {Tornambe}, A. 1978, \aap, 67, 107


\bibitem[{{Carretta} {et~al.}(2009){Carretta}, {Bragaglia}, {Gratton},
  {D'Orazi}, \& {Lucatello}}]{Carretta2009}
{Carretta}, E., {Bragaglia}, A., {Gratton}, R., {D'Orazi}, V., \& {Lucatello},
  S. 2009, \aap, 508, 695


\bibitem[{{Catelan} {et~al.}(2004){Catelan}, {Pritzl}, \&
  {Smith}}]{Catelan2004}
{Catelan}, M., {Pritzl}, B.~J., \& {Smith}, H.~A. 2004, \apjs, 154, 633


\bibitem[{{Clement} {et~al.}(2001){Clement}, {Muzzin}, {Dufton}, {Ponnampalam},
  {Wang}, {Burford}, {Richardson}, {Rosebery}, {Rowe}, \& {Hogg}}]{Clement2001}
{Clement}, C.~M., {Muzzin}, A., {Dufton}, Q., {Ponnampalam}, T., {Wang}, J.,
  {Burford}, J., {Richardson}, A., {Rosebery}, T., {Rowe}, J., \& {Hogg}, H.~S.
  2001, \aj, 122, 2587


\bibitem[{{Clement} \& {Shelton}(1997)}]{Clement1997}
{Clement}, C.~M. \& {Shelton}, I. 1997, \aj, 113, 1711


\bibitem[{{Clementini} {et~al.}(2019){Clementini}, {Ripepi}, {Molinaro},
  {Garofalo}, {Muraveva}, {Rimoldini}, {Guy}, {Jevardat de Fombelle},
  {Nienartowicz}, {Marchal}, {Audard}, {Holl}, {Leccia}, {Marconi}, {Musella},
  {Mowlavi}, {Lecoeur-Taibi}, {Eyer}, {De Ridder}, {Regibo}, {Sarro},
  {Szabados}, {Evans}, \& {Riello}}]{Clementini2019}
{Clementini}, G., {Ripepi}, V., {Molinaro}, R., {Garofalo}, A., {Muraveva}, T.,
  {Rimoldini}, L., {Guy}, L.~P., {Jevardat de Fombelle}, G., {Nienartowicz},
  K., {Marchal}, O., {Audard}, M., {Holl}, B., {Leccia}, S., {Marconi}, M.,
  {Musella}, I., {Mowlavi}, N., {Lecoeur-Taibi}, I., {Eyer}, L., {De Ridder},
  J., {Regibo}, S., {Sarro}, L.~M., {Szabados}, L., {Evans}, D.~W., \&
  {Riello}, M. 2019, \aap, 622, A60


\bibitem[{{Cohen} \& {Sarajedini}(2012)}]{CohenSara2012}
{Cohen}, R.~E. \& {Sarajedini}, A. 2012, \mnras, 419, 342


\bibitem[{{Deras} {et~al.}(2019){Deras}, {Arellano Ferro}, {L{\'a}zaro},
  {Bustos Fierro}, {Calder{\'o}n}, {Muneer}, \& {Giridhar}}]{Deras2019}
{Deras}, D., {Arellano Ferro}, A., {L{\'a}zaro}, C., {Bustos Fierro}, I.~H.,
  {Calder{\'o}n}, J.~H., {Muneer}, S., \& {Giridhar}, S. 2019, \mnras, 486,
  2791


\bibitem[{{Dworetsky}(1983)}]{Dworetsky1983}
{Dworetsky}, M.~M. 1983, \mnras, 203, 917


\bibitem[{{Ferraro} {et~al.}(1993){Ferraro}, {Clementini}, {Fusi-Pecci},
  {Vitiello}, \& {Buonanno}}]{Ferraro1993}
{Ferraro}, F.~R., {Clementini}, G., {Fusi-Pecci}, F., {Vitiello}, E., \&
  {Buonanno}, R. 1993, \mnras, 264, 273


\bibitem[{{Fourcade} \& {Laborde}(1966)}]{Fourcade1966}
{Fourcade}, C.~R. \& {Laborde}, J.~R. 1966, Atlas y Cat\'alogo de Estrellas
  Variables en C\'umulos globulares al sur de -29$^{\circ}$, Cordoba, Argentina


\bibitem[{{Gaia Collaboration} {et~al.}(2018){Gaia Collaboration}, {Brown},
  {Vallenari}, {Prusti}, {de Bruijne}, {Babusiaux}, {Bailer-Jones}, {Biermann},
  {Evans}, {Eyer}, \& et~al.}]{Gaia2018}
{Gaia Collaboration}, {Brown}, A.~G.~A., {Vallenari}, A., {Prusti}, T., {de
  Bruijne}, J.~H.~J., {Babusiaux}, C., {Bailer-Jones}, C.~A.~L., {Biermann},
  M., {Evans}, D.~W., {Eyer}, L., \& et~al. 2018, \aap, 616, A1


\bibitem[{{Goldsbury} {et~al.}(2010){Goldsbury}, {Richer}, {Anderson},
  {Dotter}, {Sarajedini}, \& {Woodley}}]{Goldsbury2010}
{Goldsbury}, R., {Richer}, H.~B., {Anderson}, J., {Dotter}, A., {Sarajedini},
  A., \& {Woodley}, K. 2010, AJ, 140, 1830


\bibitem[{{Guldenschuh} {et~al.}(2005){Guldenschuh}, {Layden}, {Wan},
  {Whiting}, {van der Bliek}, {Baca}, {Carlin}, {Freismuth}, {Mora}, {Salyk},
  {Vera}, {Verdugo}, \& {Young}}]{Guldenschuh2005}
{Guldenschuh}, K.~A., {Layden}, A.~C., {Wan}, Y., {Whiting}, A., {van der
  Bliek}, N., {Baca}, P., {Carlin}, J., {Freismuth}, T., {Mora}, M., {Salyk},
  C., {Vera}, S., {Verdugo}, M., \& {Young}, A. 2005, \pasp, 117, 721


\bibitem[{{Harris}(1996)}]{Harris1996}
{Harris}, W.~E. 1996, \aj, 112, 1487


\bibitem[{{Jurcsik} \& {Kov{\'a}cs}(1996)}]{Jurcsik1996}
{Jurcsik}, J. \& {Kov{\'a}cs}, G. 1996, \aap, 312, 111


\bibitem[{{Kov{\'a}cs} \& {Kanbur}(1998)}]{Kovacs1998}
{Kov{\'a}cs}, G. \& {Kanbur}, S.~M. 1998, \mnras, 295, 834


\bibitem[{{Kov{\'a}cs} \& {Walker}(2001)}]{Kovacs2001}
{Kov{\'a}cs}, G. \& {Walker}, A.~R. 2001, \aap, 374, 264


\bibitem[{{Kunder} {et~al.}(2013){Kunder}, {Stetson}, {Cassisi}, {Layden},
  {Bono}, {Catelan}, {Walker}, {Paredes Alvarez}, {Clem}, {Matsunaga},
  {Salaris}, {Lee}, \& {Chaboyer}}]{Kunder2013a}
{Kunder}, A., {Stetson}, P.~B., {Cassisi}, S., {Layden}, A., {Bono}, G.,
  {Catelan}, M., {Walker}, A.~R., {Paredes Alvarez}, L., {Clem}, J.~L.,
  {Matsunaga}, N., {Salaris}, M., {Lee}, J.-W., \& {Chaboyer}, B. 2013, \aj,
  146, 119


\bibitem[{{Laborde} \& {Fourcade}(1966)}]{Laborde1966}
{Laborde}, J.~R. \& {Fourcade}, C.~R. 1966, \memsai, 37, 251


\bibitem[{{Lenz} \& {Breger}(2005)}]{Lenz2005}
{Lenz}, P. \& {Breger}, M. 2005, Communications in Asteroseismology, 146, 53


\bibitem[{{Morgan} {et~al.}(2007){Morgan}, {Wahl}, \&
  {Wieckhorst}}]{Morgan2007}
{Morgan}, S.~M., {Wahl}, J.~N., \& {Wieckhorst}, R.~M. 2007, \mnras, 374, 1421


\bibitem[{{Poretti} {et~al.}(2008){Poretti}, {Clementini}, {Held}, {Greco},
  {Mateo}, {Dell'Arciprete}, {Rizzi}, {Gullieuszik}, \& {Maio}}]{Poretti2008}
{Poretti}, E., {Clementini}, G., {Held}, E.~V., {Greco}, C., {Mateo}, M.,
  {Dell'Arciprete}, L., {Rizzi}, L., {Gullieuszik}, M., \& {Maio}, M. 2008,
  \apj, 685, 947


\bibitem[{{Salaris} {et~al.}(1993){Salaris}, {Chieffi}, \& {Straniero}}]{Sal93}
{Salaris}, M., {Chieffi}, A., \& {Straniero}, O. 1993, \apj, 414, 580


\bibitem[{{Salinas} {et~al.}(2007){Salinas}, {Catelan}, {Smith}, \&
  {Pritzl}}]{Salinas2007}
{Salinas}, R., {Catelan}, M., {Smith}, H.~A., \& {Pritzl}, B.~J. 2007,
  Information Bulletin on Variable Stars, 5744


\bibitem[{{Salinas} {et~al.}(2016){Salinas}, {Contreras Ramos}, {Strader},
  {Hakala}, {Catelan}, {Peacock}, \& {Simunovic}}]{Salinas2016}
{Salinas}, R., {Contreras Ramos}, R., {Strader}, J., {Hakala}, P., {Catelan},
  M., {Peacock}, M.~B., \& {Simunovic}, M. 2016, \aj, 152, 55


\bibitem[{{Schlafly} \& {Finkbeiner}(2011)}]{Schlafly2011}
{Schlafly}, E.~F. \& {Finkbeiner}, D.~P. 2011, \apj, 737, 103


\bibitem[{{Schlegel} {et~al.}(1998){Schlegel}, {Finkbeiner}, \&
  {Davis}}]{Schlegel1998}
{Schlegel}, D.~J., {Finkbeiner}, D.~P., \& {Davis}, M. 1998, \apj, 500, 525


\bibitem[{{Stetson}(2000)}]{Stetson2000}
{Stetson}, P.~B. 2000, \pasp, 112, 925


\bibitem[{{Sturch}(1966)}]{Sturch1966}
{Sturch}, C. 1966, \apj, 143, 774


\bibitem[{{VandenBerg} {et~al.}(2014){VandenBerg}, {Bergbusch}, {Ferguson}, \&
  {Edvardsson}}]{Vandenberg2014}
{VandenBerg}, D.~A., {Bergbusch}, P.~A., {Ferguson}, J.~W., \& {Edvardsson}, B.
  2014, \apj, 794, 72


\bibitem[{{VandenBerg} {et~al.}(2013){VandenBerg}, {Brogaard}, {Leaman}, \&
  {Casagrande}}]{Vandenberg2013}
{VandenBerg}, D.~A., {Brogaard}, K., {Leaman}, R., \& {Casagrande}, L. 2013,
  \apj, 775, 134


\bibitem[{{Wehlau} \& {Demers}(1977a)}]{Wehlau1977a}
{Wehlau}, A. \& {Demers}, S. 1977a, \aap, 57, 251


\bibitem[{{Wehlau} {et~al.}(1977b){Wehlau}, {Flemming}, {Demers}, \&
  {Bartolini}}]{Wehlau1977b}
{Wehlau}, A., {Flemming}, T., {Demers}, S., \& {Bartolini}, C. 1977b,
  Information Bulletin on Variable Stars, 1361


\bibitem[{{Zinn} \& {West}(1984)}]{Zinn1984}
{Zinn}, R. \& {West}, M.~J. 1984, \apjs, 55, 45


\end{thebibliography}

\end{document}